\documentclass[runningheads]{svmult}
\usepackage{makeidx}   
\usepackage{graphicx}  
\usepackage{subeqnar}  
\usepackage{multicol}  
\usepackage{cropmark} 
\usepackage{physprbb}  
\def\apj#1{{\em Astrophys. J.} {\bf #1}}
\def\mn#1{{\em Mon. Not. R. astr. Soc.} {\bf #1}}
\def\aa#1{{\em Astron. Astrophys.} {\bf #1}}
\def\nat#1{{\em Nature} {\bf #1}}

\def\etal{{et al\/}\ }
\def\Mpc{$h^{-1}$~{\rm  Mpc}}
\def\hmpc{$h$~{\rm  Mpc$^{-1}$}}

\begin{document}

\title*{Large-Scale Structure and Dark Matter Problem}

\toctitle{Large-Scale Structure and Dark Matter Problem}
\titlerunning{Large-Scale Structure}
\author{Jaan Einasto\inst{}}
\authorrunning{Jaan Einasto}

\institute{ Tartu Observatory, EE-61602 T\~oravere, Estonia}

\maketitle              

\begin{abstract}
I review the observational data most relevant for large scale
structure.  These data determine the system of cosmological
parameters: the Hubble parameter, densities of various populations of
the Universe, parameters characterizing the power spectrum of matter,
including the biasing parameter of galaxies relative to matter.
Recent data suggest that the overall matter/energy density is
approximately equal to the critical density, and most ($0.6 - 0.7$) of
the density is in the form of cosmological term or ``dark (vacuum)
energy''.  The density of the matter is $0.3 - 0.4$ (including hot and
cold dark matter and luminous matter), the upper limit of the density
of the hot dark matter is 0.05, all in units of the critical
cosmological density.
\end{abstract}

\section{Introduction}

Recent results from the supernova cosmology project \cite{perl98},
\cite{riess98} and measurements of the angular spectrum of the cosmic
microwave background (CMB) radiation \cite{boomerang}, \cite{maxima}
have triggered a number of efforts to determine a concordant system of
cosmological parameters.  In this talk I shall use recent
observational data to discuss values of main cosmological parameters.
In addition to data on the CMB angular spectrum and supernova
cosmology project I shall use data based on the large-scale
distribution of galaxies and clusters of galaxies -- the power
spectrum, the cluster mass function etc.  In this analysis I use the
following assumptions: 1) the main constituents of the Universe are
baryonic matter, cold dark matter (CDM) with some mixture of hot dark
matter (HDM), and the dark (vacuum) energy; 2) power spectra of
galaxies and CMB radiation are determined by the initial
post-inflational power spectrum and by physical processes during the
radiation--dominated era. These processes depend on cosmological
parameters (properties of various components of the matter), and on
geometrical properties of the Universe.  In this analysis I try to
find the possible range of cosmological parameters and to show how
these are affected by the various types of data.

\section{Observed quantities and functions}

The Hubble parameter, $H_0 = 100~h$ km~s$^{-1}$~Mpc$^{-1}$, is the
observable quantity that can be estimated directly.  There exist
several methods to its estimation through the ladder of various
distance estimators from star clusters to cepheids in nearby galaxies,
through the light curves of medium-distant supernovas, and using
several physical effects (gravitational lensing, SZ-effect).
Summaries of recent determinations are given by \cite{parodi00} and
\cite{sakai00}).   I shall use here a value $h= 0.65 \pm 0.07$.

The baryon density, $\Omega_b$, can be determined most accurately
through observations of the deuterium, helium and lithium abundances
in combination with the nucleosynthesis constrains.  The best
available result is $\Omega_b h^2 = 0.019 \pm 0.002$ \cite{burles99},
\cite{burles00}. 

The total density (including vacuum energy), $\Omega_{tot} =
\Omega_m + \Omega_v$, determines the position of the first Doppler
peak of the angular spectrum of CMB temperature fluctuations; here
$\Omega_m$ and $\Omega_v$ are densities of the matter and the 
vacuum energy, respectively. Recent observations show that the
maximum of the first Doppler peak lies at $l \approx 200$
\cite{boomerang}, \cite{maxima}, \cite{seljak00}.  This indicates that
$\Omega_{tot} \approx 1$.  Since this is the theoretically preferred
value, I assume in the following that $\Omega_{tot} = 1$.

There exist a number of methods to estimate the total density of
matter (without vacuum energy), $\Omega_m = \Omega_b + \Omega_c +
\Omega_n$, where $\Omega_b$, $\Omega_c$, and $\Omega_n$ are densities
of the baryonic matter, the cold dark matter (CDM), and the hot dark
matter (HDM), respectively.  A direct method is based on the distant
supernova project, which yields (for a spatially flat universe)
$\Omega_m = 0.28 \pm 0.05$ \cite{perl98}, \cite{riess98},
\cite{gold00}.  Another method is based on X-ray data on clusters of
galaxies, which gives the fraction of gas in clusters, $f_{gas} =
\Omega_b/\Omega_m$.  If compared to the density of the baryonic matter
one gets the estimate of the total density, $\Omega_m = 0.31 \pm 0.05
(h/0.65)^{-1/3}$ \cite{mohr00}, \cite{henr00}.  A third method is
based on the geometry of the Universe.  Observations show the presence
of a dominant scale, $l_0 = 130 \pm 10$~\Mpc, in the distribution of
high-density regions \cite{beks}, \cite{e97a}, \cite{e97b}.  A similar
phenomenon is observed in the distribution of Lyman-break galaxies
\cite{bj00} at high redshift, $z \approx 3$.  We can assume that this
scale is primordial and co-moves with the expansion; in other words --
it can be used as a standard ruler.  The relation between redshift
difference and linear comoving separation depends on the density
parameter of the Universe; for a spatially flat Universe one gets a
density estimate $\Omega_m = 0.4 \pm 0.1$.  The same method was
applied for the distribution of quasars by \cite{rm00} with the result
$\Omega_m = 0.3 \pm 0.1$.  Finally, the evolution of the cluster
abundance with time also depends on the density parameter (see
\cite{b99} for a review). This method yields an estimate $\Omega_m =
0.4 \pm 0.1$ for the matter density.  The formal weighted mean of
these independent estimates is $\Omega_m = 0.32 \pm 0.03$.

Cosmological parameters enter as arguments in a number of functions
which can be determined from observations.  These functions include
the power spectrum of galaxies, the angular spectrum of temperature
fluctuations of the CMB radiation, the cluster mass and velocity
distribution.  I accept the power spectrum of galaxies according to a
summary by Einasto \etal \cite{e99a} with the addition of the recent
determination of the cluster power spectrum by Miller \& Batuski
\cite{mb00}.  The amplitude of the power spectrum can be expressed
through the $\sigma_8$ parameter, which describes the rms density
fluctuations within a sphere of radius 8~\Mpc. This parameter was
determined for the present epoch for galaxies, $(\sigma_8)_{gal} =
0.89 \pm 0.09$ \cite{e99a}.  For the CMB angular spectrum I use recent
BOOMERANG and MAXIMA I measurements \cite{boomerang}, \cite{maxima}.
For the cluster mass distribution I use determinations by Bahcall \&
Cen \cite{bc93} and Girardi et al. \cite{gir98}, see
Figures~\ref{fig2}, \ref{fig3}.

\section{Relations between cosmological parameters and observed
quantities}

I consider the following cosmological parameters: the Hubble
parameter, $h$; densities of the main constituents of the Universe:
the baryonic matter, $\Omega_b$; CDM, $\Omega_c$; HDM, $\Omega_n$; and
dark energy, $\Omega_v$ (in units of the critical cosmological
density); the index of the primordial power spectrum, $n$; the
parameter $\sigma_8$, characterizing the amplitude of the spectrum;
and the biasing parameter of the clustered matter, $b_c$. I use the
definition of the biasing parameter through the ratio of the power
spectrum of all matter to that of the clustered matter, associated
with galaxies,
$$
P_c(k) = b_c^2(k) P_m(k).
\eqno(1)
$$
Here $k$ is the wavenumber in units of $h$~Mpc$^{-1}$. In general, the
biasing parameter is a function of wavenumber $k$.  I assume that in
the linear regime of the structure evolution the biasing parameter is
constant.  Calculations show that this assumption is correct for
wavenumbers smaller than $k \approx 0.8$~\hmpc, or scales larger than
about 8~\Mpc\ \cite{e99b}.

The power spectra of matter and the angular spectra of CMB were
calculated for a set of cosmological parameters using the CMBFAST
algorithm \cite{sz96}; spectra are COBE normalized.
The cluster abundance and mass distribution functions were calculated
using the Press-Schechter  algorithm \cite{ps74} for the same set of
cosmological parameters.

Power spectra of matter and galaxies are related through the biasing
parameter.  The power spectrum is proportional to the square of the
amplitude of the density contrast. The clustered population associated
with galaxies does not include the matter in voids. If we subtract
from the density field of all matter an approximately constant density
background of void matter to get the density field of the clustered
matter, then amplitudes of absolute density fluctuations remain the
same, but amplitudes of the density contrast increase by a factor
which is equal to the ratio of mean densities of both fields, i.e. by
the fraction of matter in the clustered population, $F_c$. We obtain
\cite{e99b}
$$
b_c = 1/F_c.
\eqno(2)
$$

The possible range of the bias was determined by numerical
simulations.  During the dynamical evolution matter flows away from
low-density regions and forms filaments and clusters of galaxies.
This flow depends slightly on the density parameter of the model. The
fraction of matter in the clustered population was found by counting
particles with local density values exceeding a certain threshold
(mean density).  The present epoch of simulations was expressed
through the $\sigma_8$ parameter.  This quantity was calculated by
integrating the power spectrum of matter.  It is related to the
observed value of $(\sigma_8)_{gal}$ by the following equation:
$$
(\sigma_8)_{gal} = b_{gal} (\sigma_8)_m.
\eqno(3)
$$

\begin{figure}[ht]
\vspace*{10.8cm}
\caption{ Upper left: the fraction of matter in the clustered
population associated with galaxies as a function of $\sigma_8$ for
two LCDM models (dashed curves) and the relation between $F_{gal}$ and
$(\sigma_8)_m$ (bold solid line) defined by eq. (3). Upper
right: the biasing parameter needed to bring the amplitude $\sigma_8$
of the model into agreement with the observed $\sigma_8$ for galaxies;
for LCDM and MDM models with various matter density $\Omega_m$ and HDM
density, $\Omega_n$. Dashed box shows the range of the bias parameter
allowed by numerical simulations of the evacuation of voids. Lower
left: power spectra of LCDM models with various $\Omega_m$. Lower
right: angular spectra of CMB for LCDM and MDM models for various
$\Omega_m$.  }
\includegraphics{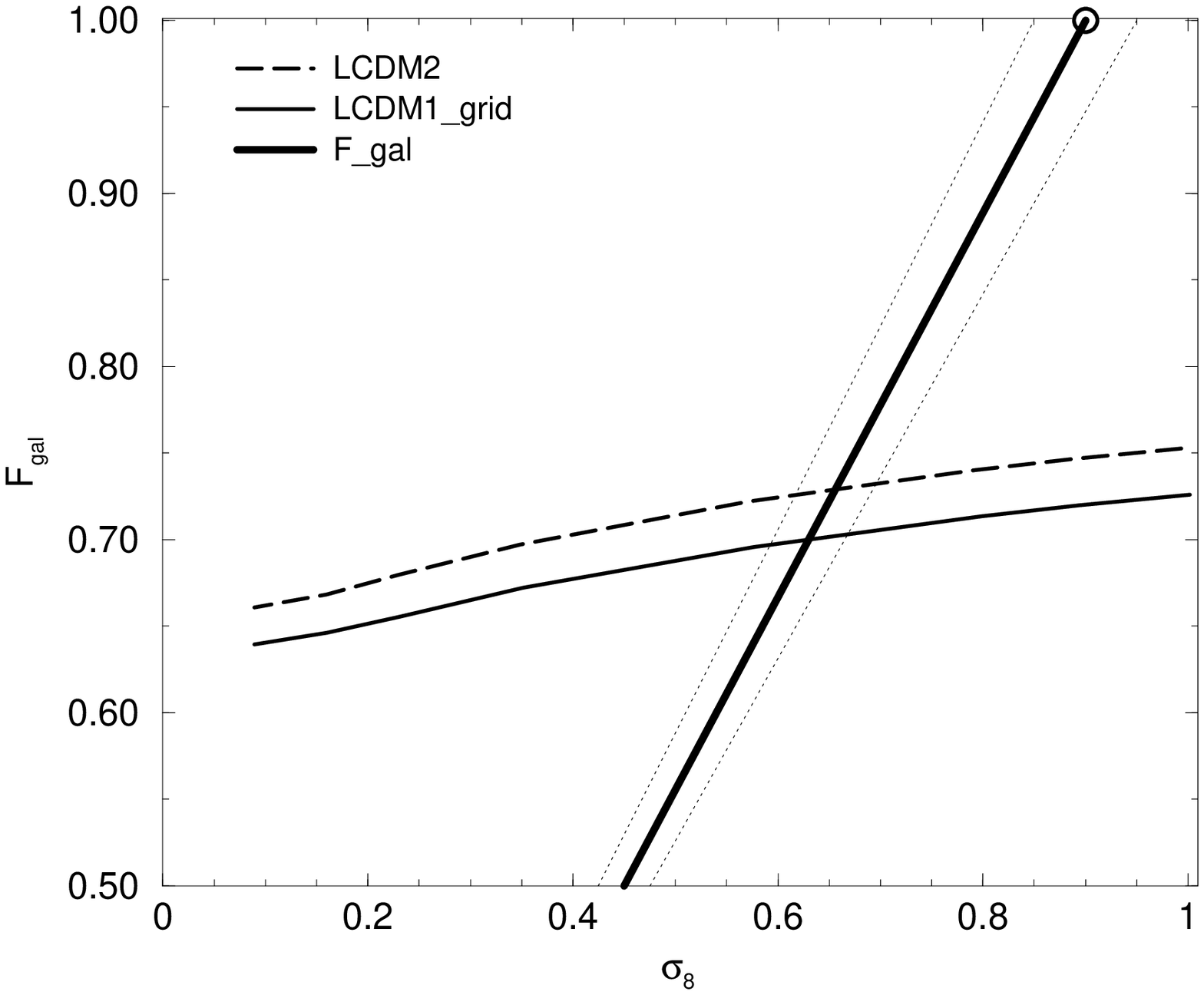} 
\includegraphics{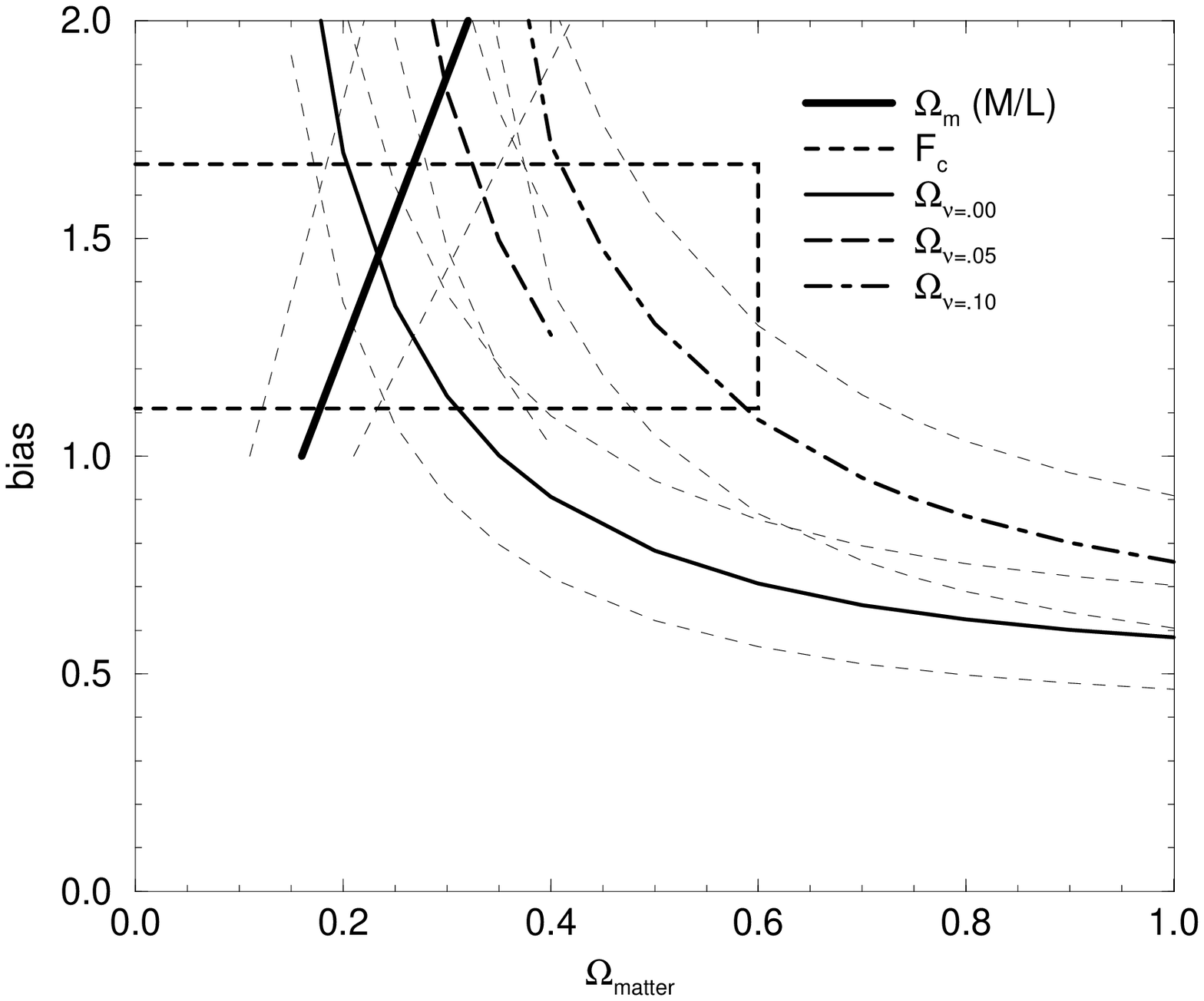}
\includegraphics{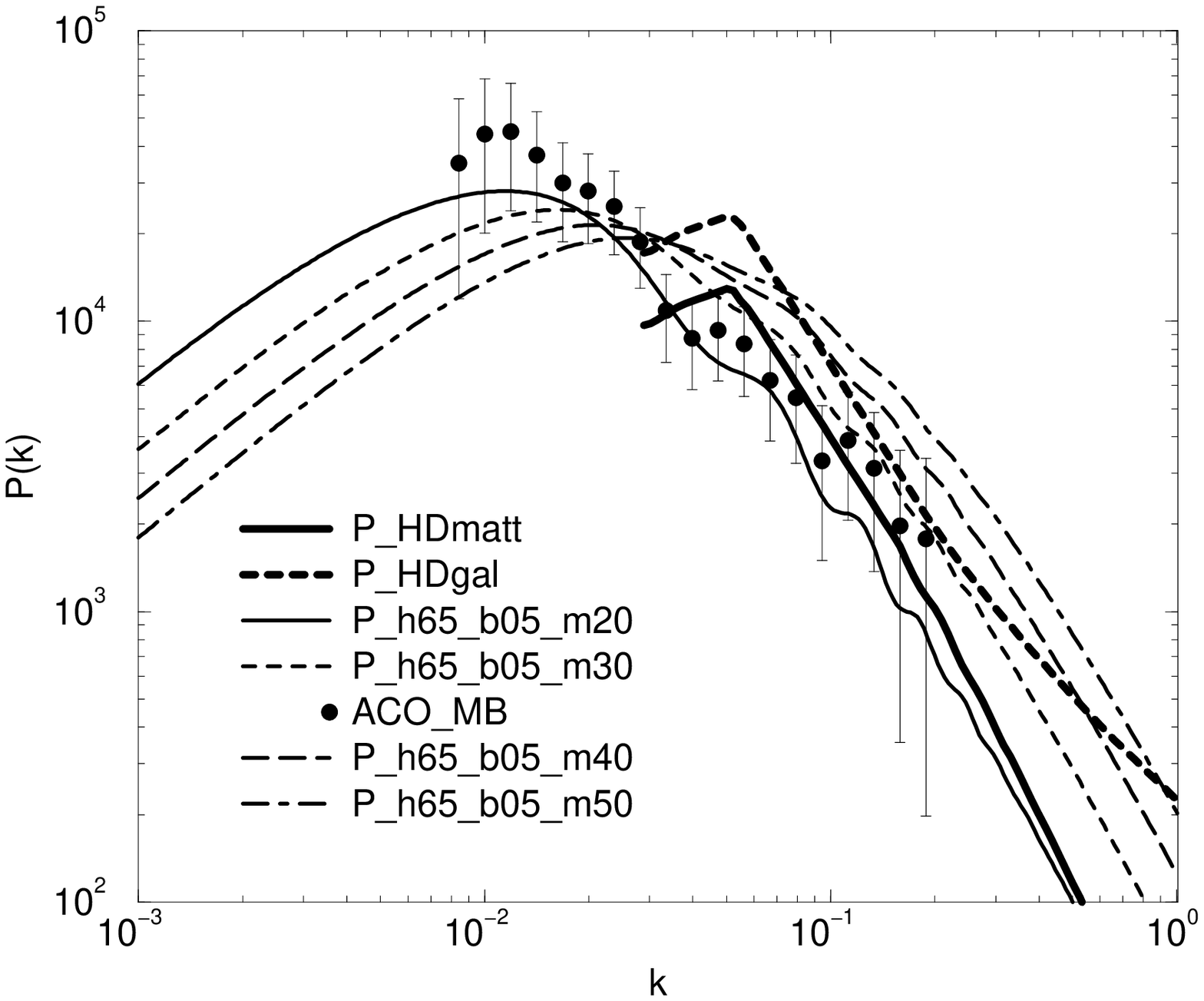} 
\includegraphics{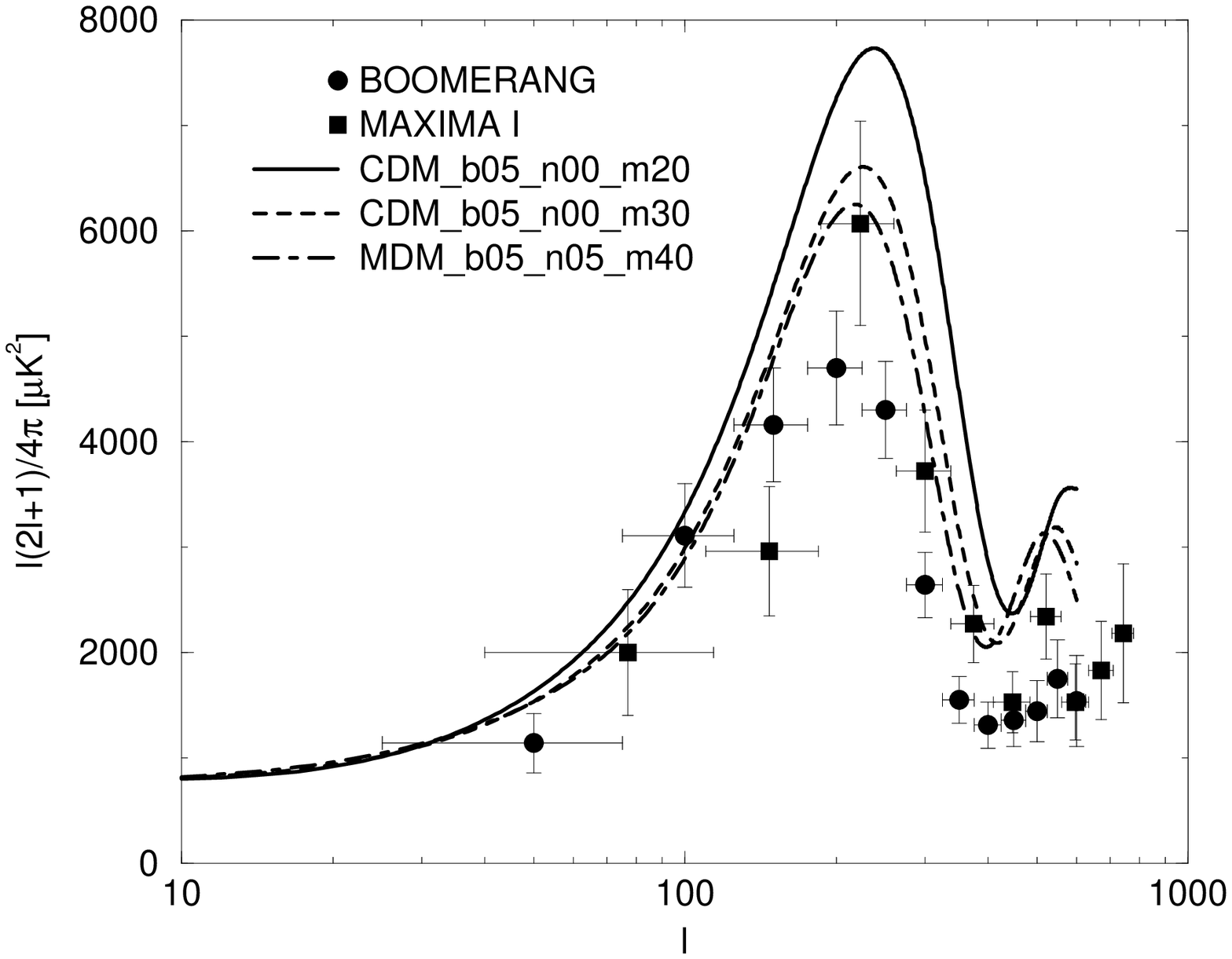}
\label{fig1}
\end{figure} 

We assume that $b_{gal} = b_c$.  For two LCDM models with density
parameter $\Omega_m \approx 0.4$ the growth of $F_{gal}$ is shown in
Fig.~1 \cite{e99b}. Using observed $(\sigma_8)_{gal}$ in
combination with relation (3) (shown in upper left panel of Fig.1 by a
bold line with error corridor), and the growth of $F_{gal}$ with epoch
(dashed curves), we get for the present epoch rms density fluctuations
of the matter $(\sigma_8)_m = 0.64 \pm 0.06$, the fraction of matter
in the clustered population, $F_{gal} = 0.70 \pm 0.09$, and the
biasing parameter $b_{gal} = 1.4 \pm 0.1$.

\section{Analysis}

The CMBFAST algorithm yields for every set of cosmological parameters
the $\sigma_8$ value for matter.  From observations we know this
parameter for galaxies, $(\sigma_8)_{gal}$.  Using eqn. (3) we can
calculate the biasing parameter $b_{gal}$, needed to bring the
theoretical power spectrum of matter into agreement with the observed
power spectrum of galaxies.  This parameter must lie in the range
allowed by numerical simulations of the evolution of structure.
Results of calculations for a range of $\Omega_m$ are shown in Fig.~1
(upper right), using a Hubble parameter of $h=0.65$, a baryon density
of $\Omega_b=0.05$, and HDM densities of $\Omega_n = 0.00,~~0.05$,  and
0.10.  The biasing parameter range shown in the Figure is larger than
expected from calculations described above; this range corresponds to
the maximum allowed range of the fraction of matter in the clustered
population expected from analytic estimates of the speed of void
evacuation.

Power spectra for LCDM models ($\Omega_n = 0$; $0.2 \leq \Omega_m \leq
0.5$) are shown in lower left panel of Fig.~1.  We see that with
increasing $\Omega_m$ the amplitude of the power spectrum on small
scales (and respective $\sigma_8$ values) increases, so that the
amplitude of the matter power spectrum exceeds for high $\Omega_m$ the
amplitude of the galaxy power spectrum. This leads to bias parameter
values $b \leq 1$.  Such values are unlikely since {\em the presence
of matter in voids always increases the amplitude of the galaxy power
spectrum relative to the matter spectrum}.  If other constraints
demand a higher matter density value, then the amplitude of the matter
power spectrum can be lowered by adding some amount of HDM. However,
supernova and cluster X-ray data exclude density values higher than
$\Omega_m \approx 0.4$; thus the possible amount of HDM is
limited. Lower right panel of Fig.~1 shows the angular spectrum of
temperature anisotropies of CMB for some density parameter values.  We
see that a low amplitude of the first Doppler peak of the CMB spectrum
prefers a higher $\Omega_m$ value: for small density values the
amplitude is too high. Thus, a certain compromise is needed to satisfy
all data.

\begin{figure}[ht]
\vspace*{5.5cm}
\caption{ Left: cluster mass distribution for LCDM models of various
density $\Omega_m$, with and without a Chung bump of amplitude
$a=0.5$. Right: cluster abundance of LCDM and MDM models of various density of
matter $\Omega_m$ and hot dark matter $\Omega_n$.  }
\includegraphics{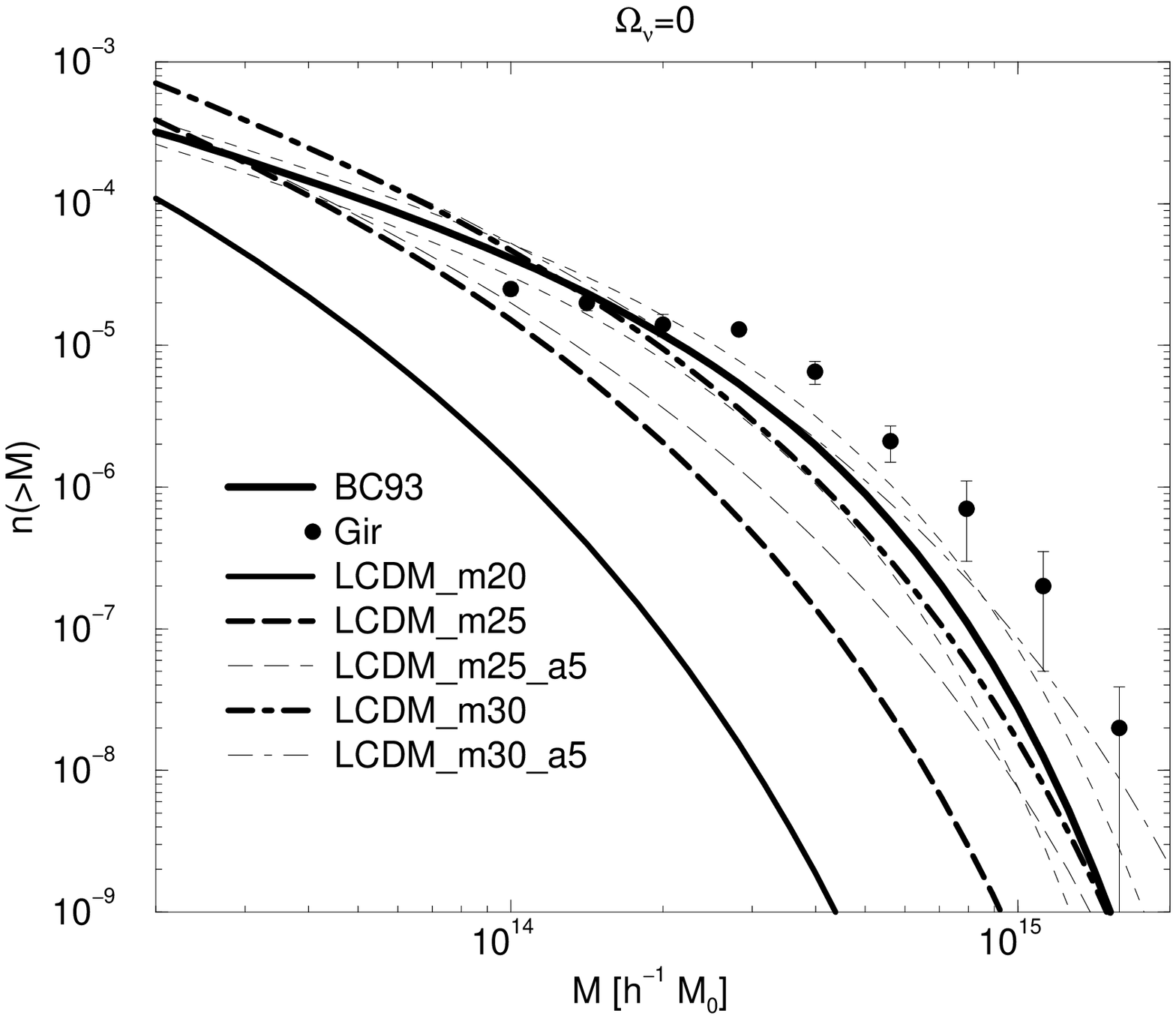} 
\includegraphics{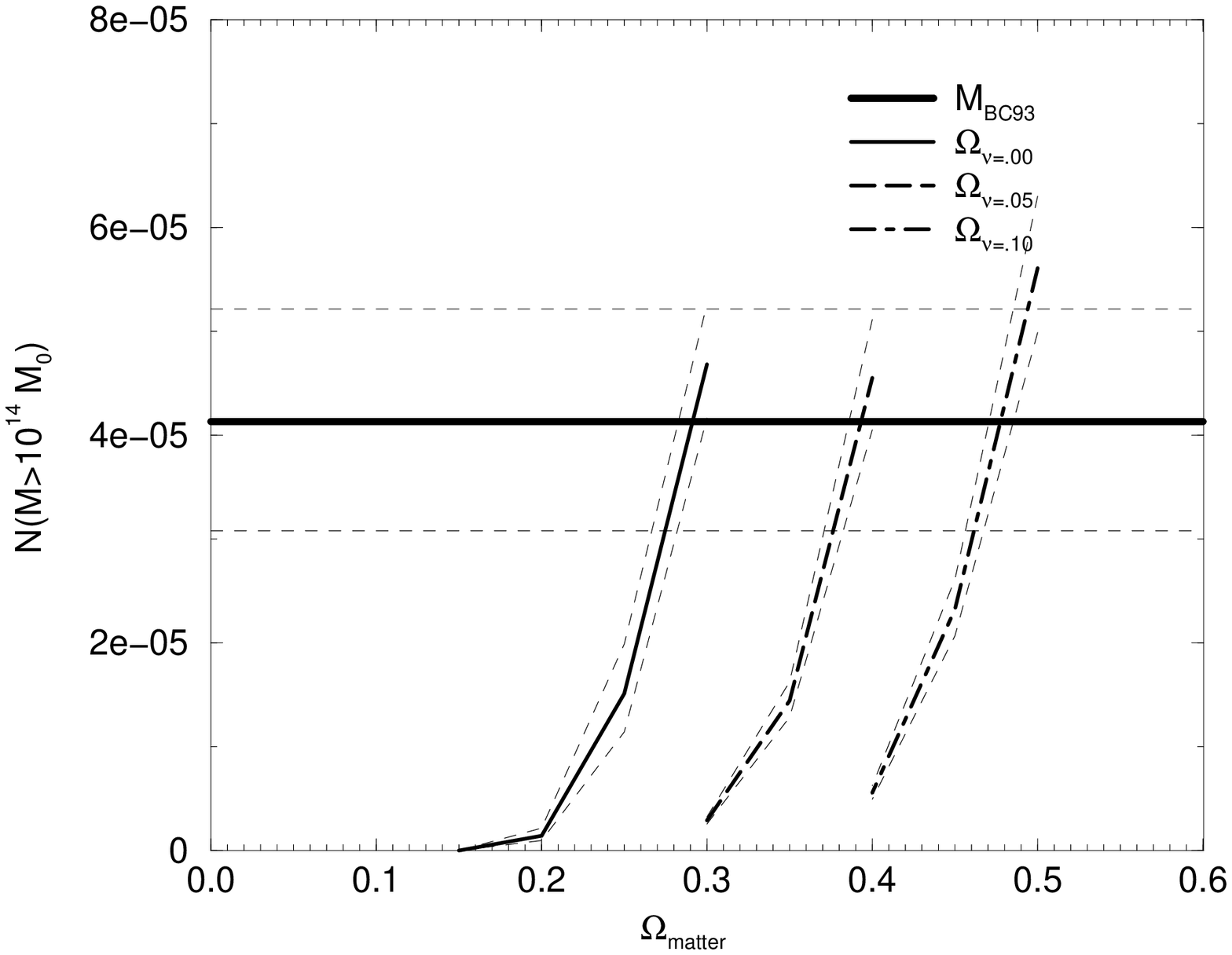}
\label{fig2}
\end{figure} 

\begin{figure}[ht]
\vspace*{10.5cm}
\caption{ Upper left: power spectra of a LCDM model with and without
Starobinsky modification.  Upper right: power spectra of MDM models
with and without Chung modification.  Lower left: cluster mass
distribution for MDM models with and without Chung modification.  Lower
right: angular power spectra of tilted MDM models with and without
Chung modification (amplitude parameter $a=0.3$).  
}
\includegraphics{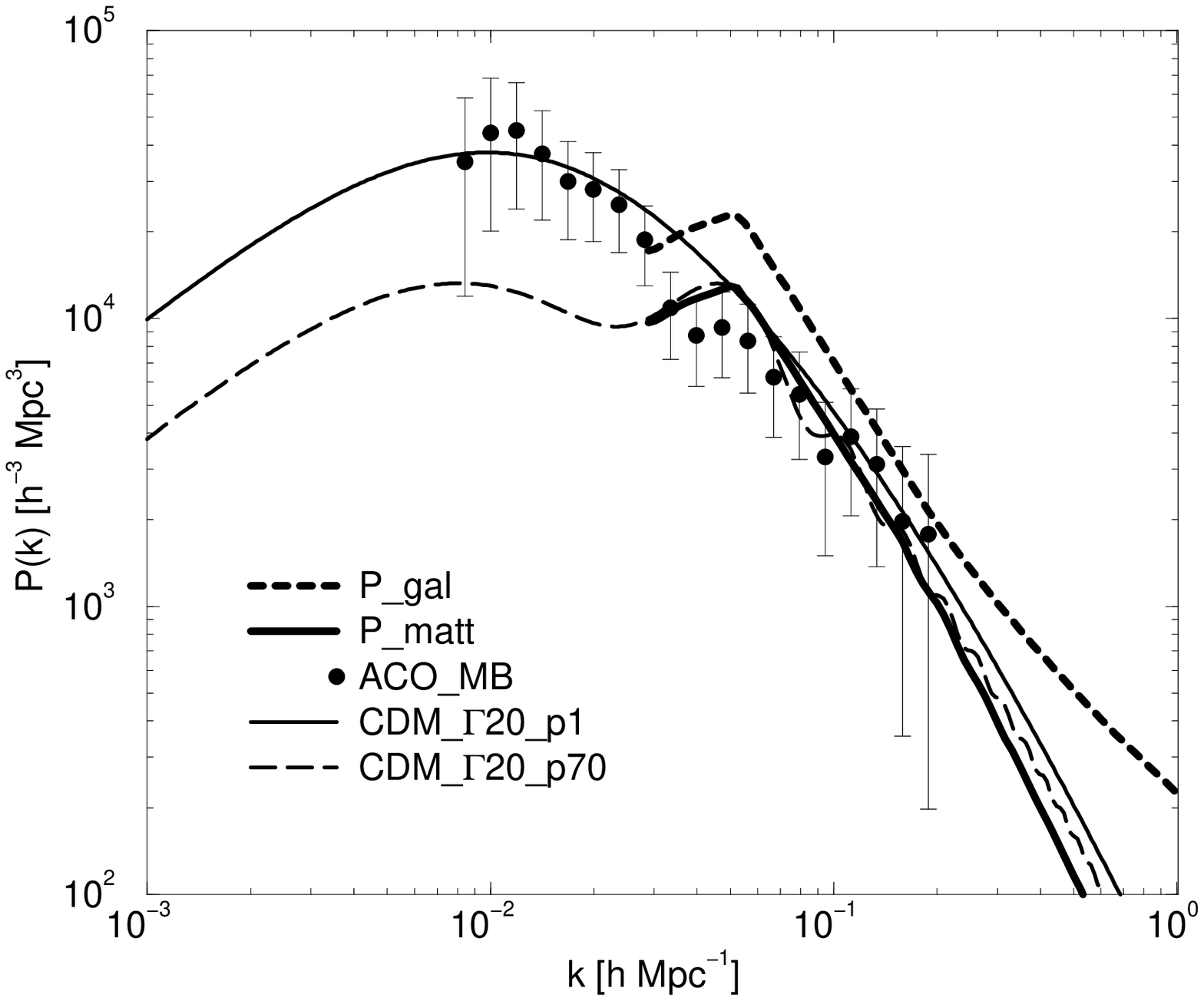} 
\includegraphics{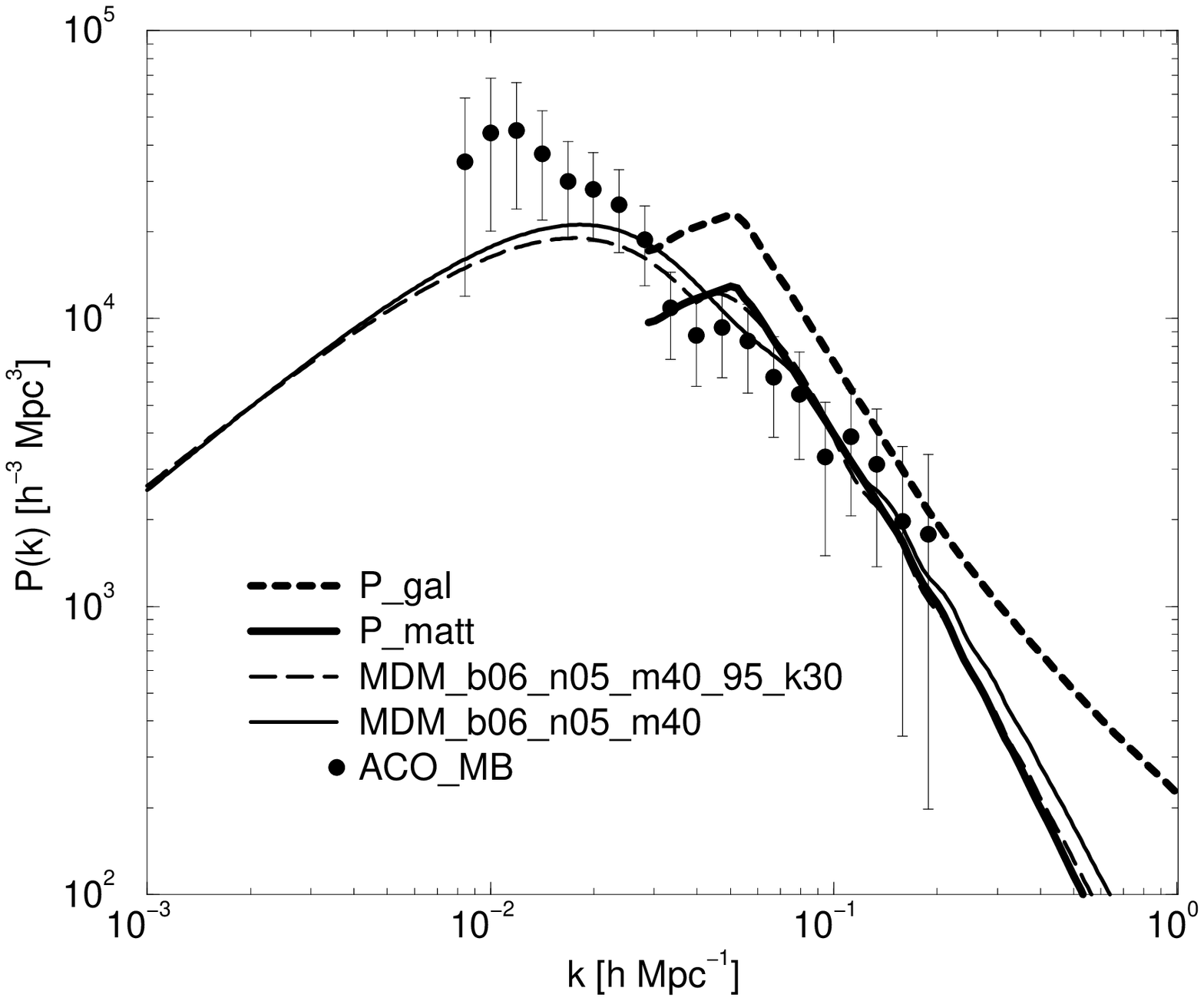}
\includegraphics{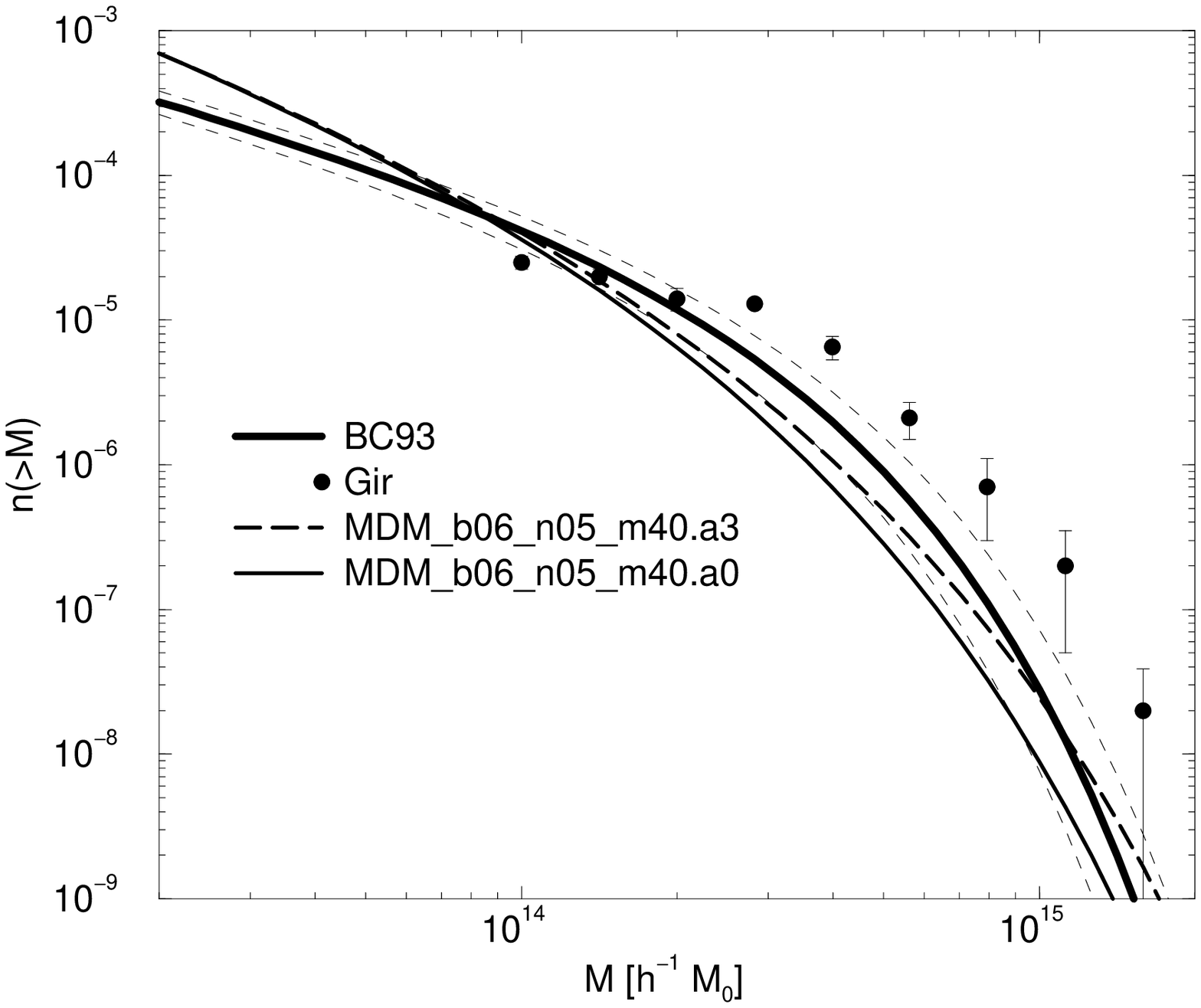} 
\includegraphics{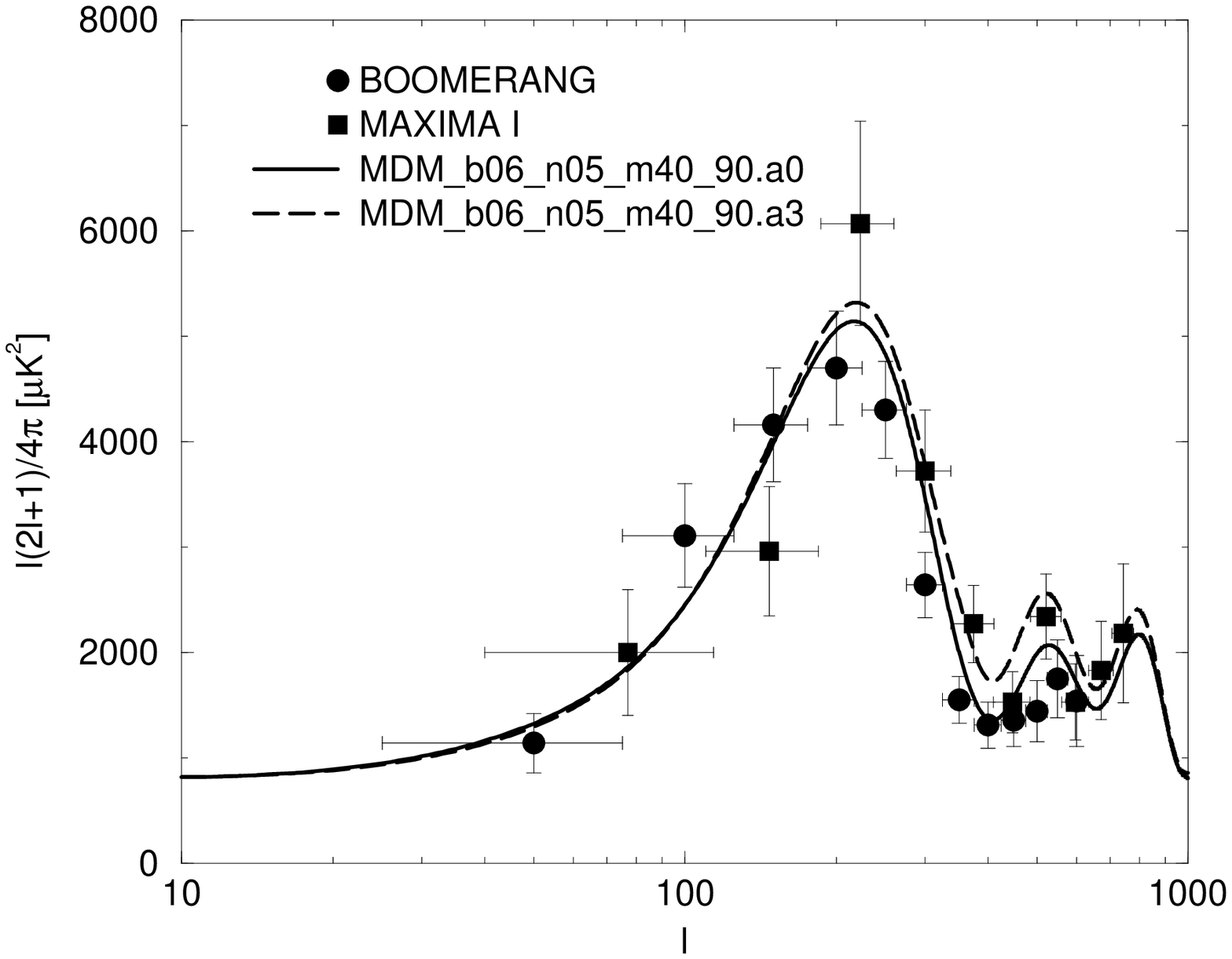}
\label{fig3}
\end{figure}

The cluster mass distribution for LCDM models with $0.2 \leq \Omega_m
\leq 0.3$ is shown in the left panel of Fig.~2.  We see that
low-density models have too low abundance of clusters over the whole
range of cluster masses.  The best agreement with the observed cluster
abundance is achieved by an LCDM model with $\Omega_m = 0.3$, in good
agreement with direct data on matter density.  In this Figure we show
also the effect of a bump in the power spectrum, which is seen in the
observed power spectrum of galaxies and clusters \cite{e99a}.  Several
modifications of the inflation scenario predict the formation of a
break or bump in the power spectrum. The influence of the break
suggested by Lesgourgues, Polarski and Starobinsky \cite{lps98} was
studied in \cite{gh00}.  Another mechanism was suggested by Chung
\etal \cite{chung}. To investigate this case we have used for the long
wavenumber end of the bump a value $k_0 = 0.04$~\hmpc, and for the
amplitude parameter $a = 0.3 - 0.8$.  Our results show that such bump
increases only the abundance of very massive clusters.  In the right
panel Fig.~2 we show the cluster abundance constraint for clusters of
masses exceeding $10^{14}$ solar masses; the curves are calculated for
LCDM and MDM models with $\Omega_n = 0.00,~~0.05,~~0.10$.  We see that
the cluster abundance criterion constrains the matter and HDM
densities in a rather narrow range.

The power spectra of LCDM models with and without the Starobinsky
break are shown in Fig.~3, upper left; these models were calculated
for the parameter $\Gamma = \Omega_m h = 0.20$.  In the case of the
spectrum with a bump we have used MDM models as reference due to the
need to decrease the amplitude of the spectrum on small scales; these
spectra are shown in Fig.~3, upper right.  Power spectra are compared
with observed galaxy power spectrum \cite{e99a}  and the new
cluster power spectrum by Miller \& Batuski \cite{mb00}, reduced to the
amplitude of the galaxy power spectrum.  We also show the matter power
spectrum based on a biasing factor $b_c = 1.3$ \cite{e99b}. 
We see that the Starobinsky model reproduces well the matter power
spectrum on small and intermediate scales, but not the new data by
Miller \& Batuski.  The modification by Chung \etal \cite{chung} with
amplitude parameter $a=0.3$ fits well all observational data.  The
cluster mass distribution for the Chung model is shown in lower left
panel of Fig.~3, and the angular spectrum of CMB temperature
fluctuations in lower right panel.  In order to fit simultaneously the
galaxy power spectrum and the CMB angular spectrum we have used a
tilted MDM model with parameters $n=0.90$, $\Omega_b = 0.06$,
$\Omega_n = 0.05$, and $\Omega_m =0.4$.

\section{Discussion}

BOOMERANG and MAXIMA I data have been used in a number of studies to
determine cosmological parameters \cite{bz00}, \cite{cen00},
\cite{boomerang}, \cite{maxima}, \cite{kinney00}, \cite{tz00},
\cite{wsp00}.  In addition to CMB data various other observational
data have been used.  In general, the agreement between various
determinations is good; however, some parameters differ.  For
instance, \cite{tz00} interpreted new CMB data in terms of a baryon
fraction higher than expected from the nucleosynthesis constraint,
$h^2 \Omega_b = 0.03$, and a relatively high matter density, $h^2
\Omega_m = 0.33$.  On the other hand, velocity data suggest a
relatively high amplitude of the power spectrum, $\sigma_8
\Omega_m^{0.6} = 0.54$, which in combination with distant supernova
data yields $\Omega_m = 0.28 \pm 0.10$ and $\sigma_8 = 1.17 \pm 0.2$
\cite{bz00}.

Our analysis has shown that a high value of the density of matter,
$\Omega_m >0.4$, and high amplitude of the matter power spectrum,
$\sigma_8 >1$, are difficult to explain in terms of the supernova and
cluster abundance data, and the observed amplitude of the galaxy power
spectrum with reasonable bias limits.  This conflict can be avoided
using a tilted initial power spectrum, and a MDM model with a moderate
fraction of HDM, as discussed above. The best models suggested so far
have $0.3 \leq \Omega_m \leq 0.4$, $0.90 \leq n \leq 0.95$, $0.60 \leq
h \leq 0.70$, $\Omega_n \leq 0.05$.  Matter densities are constrained to
$\ge 0.3$ by cluster abundances, and to $\le 0.4$ by all existing
matter density estimates. This upper limit of the matter density, in
combination with the cluster abundance and amplitude of the power
spectrum, yields an upper limit to the density of the hot dark matter.
We can consider this range of cosmological parameters as compatible
with all constraints.  This set of cosmological parameters is
surprisingly close to the set suggested by Ostriker \& Steinhardt
\cite{os95}.   Now it is supported by much more accurate observational data.

A considerably lower value of matter density, $\Omega_m = 0.16$, was
suggested by Bahcall \etal \cite{b00} from the observed value of $M/L$ for
galaxies and clusters of galaxies of various richness. Upper right
panel of Fig.~1 shows this constraint for various fractions of matter
in voids and respective bias parameter values. The reason for the
deviation of this matter density determination from the rest is not
clear, and we have not used it in the present analysis.

\section*{Acknowledgments}

I thank M. Einasto, M. Gramann, V. M\"uller, E. Saar, A. Starobinsky, 
and E. Tago for fruitful collaboration and permission to use our joint
results in this talk, Joe Silk for discussion, and H. Andernach for
help in improving the style.  This study was supported by the Estonian
Science Foundation grant 2625.

\end{document}